\documentclass[a4paper,3p,twocolumn,preprint]{elsarticle}

 \journal{Physica D}

\usepackage{amsmath}
\usepackage{amssymb}
\usepackage{graphicx}
\usepackage{bm}
\usepackage{epstopdf}

\newcommand{\la}{\lambda}
\newcommand{\ga}{\gamma}
\newcommand{\al}{\alpha}
\newcommand{\om}{\omega}
\newcommand{\K}{\mathrm{K}}

\newcommand{\sn}{\mathrm{sn}}
\newcommand{\prt}{\partial}

\begin{document}

\begin{frontmatter}

\title{
Whitham theory for perturbed Korteweg-de Vries equation}

\author{A.M.~Kamchatnov}
\ead{kamch@isan.troitsk.ru}
\address{
Institute of Spectroscopy, Russian Academy of Sciences, Troitsk,
Moscow, 142190, Russia }

\date{\today}

\begin{abstract}
Original Whitham's method of derivation of modulation equations is applied to systems
whose dynamics is described by a perturbed Korteweg-de Vries equation. Two situations
are distinguished: (i) the perturbation leads to appearance of right-hand sides in the
modulation equations so that they become non-uniform; (ii) the perturbation leads to
modification of the matrix of Whitham velocities. General form of Whitham modulation
equations is obtained for each case. The essential difference between them is illustrated
by an example of so-called `generalized Korteweg-de Vries equation'. Method of finding
steady-state solutions of perturbed Whitham equations in the case of dissipative
perturbations is considered.
\end{abstract}


\end{frontmatter}


{\it Dedicated to fiftieth anniversary of publication of Whitham's `the Paper'.}

\section{Introduction}

In his seminal paper \cite{whitham-65} Whitham introduced into nonlinear wave theory several
fundamental ideas which formed the basis for development of a vast theory called now {\it Whitham theory}.
First, he generalized the idea of slow evolution
of envelopes of linear harmonic wave trains (`wave packets') to description of evolution of nonlinear
modulated wave trains whose dynamics is governed by nonlinear wave equations. This idea implies that
in the problem under consideration there are two different scales of space and time: the field
variables $u(x,t)$ of the nonlinear `carrier' wave oscillate at the scales of wavelength $L$ and period $T$,
whereas such parameters of the wave as, e.g., wavelength $L$, amplitude $a$, phase velocity $V$,
etc., change slowly at the space scale $x\gg L$ and the time scale $t\gg T$. This leads to the
second idea of averaging of the conservation laws of the evolution equation over fast local
oscillations analogously to the Krylov-Bogoliubov averaging technique developed in the theory of
nonlinear vibrations. However, in contrast with dynamical time-dependent systems, now the field
variables depend on time and one (or more) space coordinates and, as a result of averaging of
conservation laws, Whitham obtained the system of first order partial differential equations
now called {\it Whitham equations}. Whitham compared this approach with transition from
`microscopic' description of gas dynamics to averaged hydrodynamic description (``Indeed, the
present work is in much the same spirit as the derivation of continuum fluid mechanics from
kinetic theory.'') and this suggested the third idea of application of the averaged equations to
description of such physical phenomena as, for example, water undular .bores and collisionless shocks in plasma.
At last, as the forth idea, Whitham supposed that his modulation equations can be transformed,
by analogy with compressible fluid dynamics, to the diagonal Riemann form and he realized this
idea by means of very skillful calculations for the case of modulated nonlinear `cnoidal' wave
whose evolution is governed by the Korteweg-de Vries (KdV) equation. Thus, Whitham formulated in
\cite{whitham-65} the general method for studying modulated nonlinear waves and illustrated
fruitfulness of his approach by important nontrivial examples. Richness of ideas introduced in \cite{whitham-65}
has been spectacularly confirmed by impressive development of the Whitham theory in the past 50 years.

The first important contribution into the Whitham theory after appearance of the paper \cite{whitham-65}
was done by Gurevich and Pitaevskii \cite{gp-73} who showed that a collisionless shock (now
commonly called {\it dispersive shock wave (DSW)}) described by the KdV equation can be represented as
an expanding oscillating structure which can be approximated by a modulated cnoidal wave
whose evolution is governed by the Whitham equations. At one its edge the dispersive shock approaches
to a soliton train and at the opposite edge it tends to a small amplitude harmonic wave.
Gurevich and Pitaevskii studied self-similar solutions of the Whitham equations for a typical
examples of evolutions of an initial step-like distribution and of the general wave breaking situation.

Analytical theory developed by Gurevich and Pitaevskii was based on a specific diagonal form of modulation
equations obtained by Whitham for modulated KdV cnoidal waves. However, such a form for other nonlinear
wave equations was not known and it was not easy to find it by the direct method used by Whitham.
Actually, as it became clear later, Whitham had succeeded in finding the Riemann invariants for the KdV
equation case because this equation
belongs to a very special class of so-called `completely integrable equations' whose solutions can be
found by the inverse scattering transform (IST) method discovered independently of the Whitham theory
\cite{ggkm-67,lax-68,zs-71}. It turned out that this method generalized on quasi-periodic situations
\cite{lax-74,novikov-74} yields quasi-periodic solutions of the KdV equation which are
parameterized directly by Riemann invariants having in this case very simple mathematical meaning:
they are the edge points of gaps in the spectrum of the linear (Schr\"odinger) equation related
with the KdV equation in the IST method. As a result, multi-phase averaging method for the quasi-periodic
solutions of the KdV equation was developed by Flaschka, Forest and McLaughlin \cite{ffml-80} where
the Whitham equations were derived in Riemann diagonal form for $2N+1$ dependent variables (Riemann invariants),
$N$ being the number of phases ($N=1$ for the simplest cnoidal wave case considered by Whitham in \cite{whitham-65}).
The general method of derivation of the Whitham equations for a wide class of completely integrable
equations was suggested by Krichever \cite{krichever-88}. The generalized hodograph method of integration
of diagonal Whitham equations was developed by Tsarev \cite{tsarev-90}. This progress in mathematical theory
of integrability of nonlinear wave equations and of corresponding Whitham modulation equations has led to
a number of applications to physical problems related with formation of DSWs and deeper understanding of
qualitative properties of this phenomenon. At the same time, it became quite desirable to extend the
theory of DSWs on situations often met in physical applications when wave motion is not described
by the completely integrable equations.

From physical point of view, it seems clear that the phenomenon of formation of DSWs is related
with effects of dispersion in nonlinear wave systems and is not conditioned by complete integrability of
corresponding evolution equations. Actually, the Whitham theory was developed in \cite{whitham-65} in very
general setting under supposition of existence of periodic solutions of evolution equations and only
application of this theory to the KdV cnoidal wave was related implicitly in this paper with the complete integrability of
the KdV equation. Therefore for treatment of DSWs in general situation one should resort to analysis of
Whitham equations in a non-diagonal form when they do not have Riemann invariants and cannot be integrated
by Tsarev's generalized hodograph method. Such an analysis was done by El \cite{el-05} in an important
particular case of evolution of an initial step-like pulse whose dynamics is governed by non-dissipative
nonlinear wave equations. This ingenious method has found a number of interesting applications in which
the problem can be reduced to the study of evolution of step-like pulses.

Another typical situation appears when the evolution equation differs little from an integrable one.
For example, such a difference can appear due to small dissipation effects or weak non-uniformity of
the medium through which the wave propagates. As was indicated already by Whitham in \cite{whitham-65},
these effects lead to modulation of nonlinear periodic waves and can be considered in framework of the
averaged modulation equations. In such situations, the Whitham modulation equations can be modified
by perturbations in two possible ways: {(i)} the equations for Riemann invariants $\la_i(x,t)$ of
unperturbed equations have now `right-hand sides'
depending on $\la_i$, that is these equations become non-uniform; {(ii)} the additional terms caused
by perturbations contribute to fluxes of the conserved quantities leading to appearance in the Whitham equations of
terms proportional to the derivatives $\prt \la_j/\prt x$, that is the matrix of `velocities'
becomes non-diagonal and $\la_i$ are not Riemann invariants anymore, but the non-diagonal terms as well as
corrections to the diagonal ones are small and the modulation equations remain uniform. Of course, one
can imagine situations when both types of corrections appear in perturbed Whitham equations.

So far, mainly the first type of corrections has been considered. Physically, such corrections appear very naturally when
irreversible processes are taken into account. The Whitham averaging method for systems with
small dissipation was developed by Jimenez and Whitham \cite{jw-76} in general form without transition
to Riemann invariants. Whitham equations for $N$-phase KdV wave trains in presence of small perturbations
were derived in \cite{fml-84}, however in a form not convenient enough for applications. More practical
and instructive example of one-phase modulated KdV wave trains with account of small
Burgers viscosity was considered by Gurevich and Pitaevskii \cite{gp-87} and by Avilov, Krichever
and Novikov \cite{akn-87} (earlier the steady-state solution of this problem had been studied
by Johnson \cite{johnson-70} by a direct perturbation technique).
They derived the Whitham equations for the Riemann invariants $\la_i$ of
unperturbed problem and showed that small Burgers viscosity results in non-zero right-hand sides
of Whitham equations which provide additional contribution into evolution of these
modulation parameters $\la_i$. The analysis presented there showed that although the perturbation is
small compared with the main terms in the KdV equation, this does not mean that its contribution into
evolution of the Riemann invariants is also small. Indeed, in this case the perturbation should be
compared with a small parameter which characterizes `slowness' of modulation rather than with terms
which determine fast oscillations of the cnoidal wave. If the cnoidal wave is not modulated at all,
then dissipative terms make the only contribution into changes of the Riemann invariants and
determine slow evolution of a uniform cnoidal wave (see, e.g., \cite{akkb-03,aaa-08}). Effects of non-local damping
were considered by Gurevich and Pitaevskii in \cite{gp-91} and more general forms of local dissipation were considered
by Myint and Grimshaw in \cite{mg-95}. Quite general approach applicable to the Ablowitz-Kaup-Newell-Segur (AKNS) class
\cite{akns-74} of completely integrable equations was developed by the author for non-perturbed
\cite{kamch-94} and perturbed \cite{kamch-04} cases. In combination with simplified version \cite{kamch-90}
of the finite-gap integration method which yields the periodic solutions in a `real' form not-constrained
by any additional `reality conditions', this approach turned out to be quite effective and it has found
several non-trivial applications including propagation of KdV wave trains through a non-uniform medium
(see, e.g., \cite{egk-05,egk-07,egt-12,lpk-12}).

However, the approach described above is not applicable to situations when perturbations change the matrix of
Whitham velocities although such situations are encountered quite often. For example, if in the perturbed KdV
equation
\begin{equation}\label{eq1}
    u_t+6uu_x+u_{xxx}=R[u]
\end{equation}
the perturbation term has the form $R[u]=\epsilon F'(u)u_x$, $(\epsilon\ll1)$, then the general formulae obtained
in \cite{mg-95,kamch-04} lead to vanishing right-hand sides in the `perturbed' Whitham equations what means that
such a perturbation belongs to the type {(ii)} and a different perturbation scheme should be developed for finding
the corrected matrix of
Whitham velocities. This difference between two types of perturbations is clearly illustrated by a simple example
of perturbation $R[u]=\mathrm{const}\cdot u^2u_x$ when (\ref{eq1}) reduces to the so-called Gardner equation. This
equation is also completely integrable, the corresponding Riemann invariants and Whitham equations can
be obtained without any approximations (see \cite{pavlov-94,pre-2012}), and they do not reduce to
appearance of the right-hand sides in Whitham equations in the KdV limit of the Gardner equation.

Thus, we arrive at the problem of derivation of the approximate Whitham equations for two different situations
when either the Whitham equations acquire the right-hand side terms, or the Whitham velocities are modified by perturbations.
Here we shall confine ourselves to a simple example of
the KdV equation (\ref{eq1}) under supposition that the perturbation term is small,
\begin{equation}\label{eq2}
    |R[u]|\ll \mathrm{min}\{u^2/L,\,|u|/L^3\}.
\end{equation}
Hence, we can approximate locally the solution of (\ref{eq1}) by the cnoidal wave solution of unperturbed
KdV equation and apply the original method of Whitham \cite{whitham-65} to this more general situation.
In the next section we shall illustrate the method by its application to the already studied earlier situation
of perturbations of type (i) and then generalize it to perturbations of type (ii). In section 3 we shall
show that a specific structure of perturbation terms leads to a simple method of finding the steady-state
solutions of the Whitham equations. We conclude by the remark that the direct Whitham approach to obtaining
the modulation equations is effective enough and it can be successfully used for studying quite complicated
nonlinear wave problems.

\section{Perturbed Whitham equations}

Traveling wave solution of the unperturbed KdV equation
\begin{equation}\label{eq3}
  u_t+6uu_x+u_{xxx}=0
\end{equation}
is obtained by means of a simple
substitution $u=u(\xi),\,\xi=x-Vt$, so that after two obvious integrations we get the ordinary
differential equation
\begin{equation}\label{eq4}
  \tfrac12u_{\xi}^2=f(u),
\end{equation}
where $f(u)$ is a third degree polynomial
$(\alpha\ge\beta\ge\gamma)$,
\begin{equation}\label{eq5}
\begin{split}
  f(u)&=-A+Bu+\tfrac12Vu^2-u^3\\
  &=-(u-\alpha)(u-\beta)(u-\gamma),
  \end{split}
\end{equation}
where $V$, $A$, and $B$ are the integration constants related  with the zeros $\alpha,\,\beta,\,\gamma$
of the polynomial $f(u)$ by the formulae
\begin{equation}\label{eq6}
\begin{split}
 &V=2(\alpha+\beta+\gamma),\quad
 A=-\alpha\beta\gamma, \\
 &B=-(\alpha\beta+\beta\gamma+\gamma\alpha).
 \end{split}
\end{equation}
In a standard way the solution of Eq.~(\ref{eq4})
can be expressed in terms of Jacobi elliptic sinus function
\begin{equation}\label{eq7}
  u(x,t)=\alpha-(\alpha-\beta)\,\sn^2(\sqrt{(\alpha-\gamma)/2}\,\xi,m),
\end{equation}
where the parameter $m$ is equal to
\begin{equation}\label{eq8}
 m=\frac{\alpha-\beta}{\alpha-\gamma}.
\end{equation}
This is a periodic solution of the KdV equation (\ref{eq3}) and its wavelength is given by the formula
\begin{equation}\label{eq9}
\begin{split}
  L&=\frac1k=\int_0^Ld\xi=\oint \frac{du}{u_{\xi}}\\
  &=\frac1{\sqrt{2}}\oint \frac{du}{\sqrt{f(u)}}=2\sqrt{\frac2{\alpha-\gamma}}\,K(m),
  \end{split}
\end{equation}
where $K(m)$ is the complete elliptic integral of the first kind. We have introduced in (\ref{eq9})
the wavenumber $k=1/L$ and hence at a given point $x$ the wave oscillates with the frequency $\om=kV$,
so that the solution (\ref{eq7}) depends on the phase $\theta=kx-\om t$.

In a modulated wave the parameters $V$, $A$, $B$, or, equivalently, $\al$, $\beta$, $\ga$, become
slow functions of $x$ and $t$, that is they change little in one wavelength $L$ and one period $T\sim1/\om$.
However, according to Whitham, the solution of the (perturbed or unperturbed) KdV equation can be
approximated locally by the expression (\ref{eq7})
where $\al(x,t),\,\beta(x,t),\,\ga(x,t)$ are considered now as slow functions of $x$ and $t$
and the phase $\theta=kx-\om t$ is replaced now by a general dependence
$\theta(x,t)$. Then the wavenumber and the frequency are defined as
\begin{equation}\label{eq11}
    k=\theta_x,\quad \om=-\theta_t,
\end{equation}
and, hence, they must satisfy the compatibility condition $k_t+\om_x=0$ which has the meaning
of conservation of `number of waves' \cite{whitham-65}. In a slowly modulated wave both $k$ and $\om=kV$
are expressed in terms of the slow parameters $\al,\,\beta,\,\ga$ (see Eqs.~(\ref{eq6}) and (\ref{eq9}))
so that we arrive at the equation for these parameters,
\begin{equation}\label{eq12}
    k_t+(kV)_x=0.
\end{equation}
All that is applied to any modulated KdV wave train and the modulation can be caused either by a non-uniform
initial condition of by a perturbation term in (\ref{eq1}). Whitham discussed in \cite{whitham-65} the first
situation only and we wish here to generalize his approach to perturbed KdV equations. As was indicated in Introduction,
we have to distinguish in this case two different situations which, as we shall see, can be formulated more
precisely as follows: (i) neither $R$ nor $uR$ are space derivatives,
(ii) $R$ and/or $uR$ can be represented as space derivatives of other functions
(say, $R=\mathcal{Q}_{1,x}$ and/or $uR=\mathcal{Q}_{2,x}$).
We shall call the first situation as a {\it non-gradient perturbation} and the second one as a {\it gradient perturbation},
and we shall begin with discussion of the non-gradient perturbations.

\subsection{Whitham equations for the case of non-gradient perturbations}

In addition to (\ref{eq12}), we need two more equations for three parameters
$\al(x,t),\,\beta(x,t),\,\ga(x,t)$ and, following Whitham \cite{whitham-65}, we assume
that they can be obtained by means of averaging the conservation laws of the KdV equation (\ref{eq1}),
\begin{equation}\label{eq13}
  \begin{split}
&  u_t+(3u^2+u_{xx})_x=R,\\
&  (\tfrac12u^2)_t+(2u^3+uu_{xx}-\tfrac12u_x^2)_x=uR.
  \end{split}
\end{equation}
Averaging is defined as taking a mean value of an expression $\mathcal{P}$ along the wavelength,
\begin{equation}\label{eq14}
    \langle \mathcal{P}\rangle=\frac1L\int_0^L \mathcal{P}dx=k\oint \mathcal{P}\frac{du}{u_{x}}
    =\frac{k}{\sqrt{2}}\oint\frac{\mathcal{P}du}{\sqrt{f(u)}},
\end{equation}
where integration is taken over the whole cycle of oscillation of $u$.
Whitham averaged densities and fluxes of the conservation laws (\ref{eq13}) (with $R=0$) and obtained
two additional equations for the slow variables. If we average the right-hand sides of Eqs.~(\ref{eq13})
according to the rule (\ref{eq14}) then we obtain the perturbed Whitham equations with
the right-hand sides. The averaged conservation laws (\ref{eq13}) take the form
\begin{equation}\label{eq16}
  \begin{split}
&  \langle u\rangle_t+\langle 3u^2+u_{xx}\rangle_x=\langle R \rangle,\\
&  \langle \tfrac12u^2\rangle_t+\langle 2u^3+uu_{xx}-\tfrac12u_x^2\rangle_x=\langle uR \rangle,
  \end{split}
\end{equation}
and the condition that $R$ and $uR$ are not $x$-derivatives yields, generally speaking,
non-zero right-hand sides in these equations.
The derivatives $u_{x}^2,\,u_{xx}$ can be excluded with the use of Eqs.~(\ref{eq4}), (\ref{eq5}):
$u_x^2=2f(u)$, $u_{xx}=f'(u)=B+Vu-3u^2$, and, as Whitham indicated, it is convenient to represent the
averaged expressions in terms of a single `action function'
\begin{equation}\label{eq17}
\begin{split}
  &W(A,B,V)=-\sqrt{2}\oint\sqrt{f(u)}\,du\\
  &=-\sqrt{2}\oint\sqrt{-A+Bu+\tfrac12Vu^2-u^3}\,du,
  \end{split}
\end{equation}
and its derivatives with respect to $A,\,B,\,V$. Indeed, the wavelength (\ref{eq9}) can be written as
\begin{equation}\label{eq18}
\begin{split}
  L=\frac1k=\frac{1}{\sqrt{2}}\oint \frac{du}{\sqrt{f(u)}}=\frac{\prt W}{\prt A}\equiv W_A
  \end{split}
\end{equation}
and the necessary averages are given by
\begin{equation}\label{eq19}
  \begin{split}
&  \langle{u}\rangle=\frac{k}{\sqrt{2}}\oint \frac{udu}{\sqrt{f(u)}}=-kW_B,\\
&  \langle{\tfrac12u^2}\rangle=\frac{k}{\sqrt{2}}\oint\frac{(u^2/2)du}{\sqrt{f(u)}}=-kW_V.
  \end{split}
\end{equation}
As a result we obtain the averaged conservation laws
\begin{equation}\nonumber
    \begin{split}
    &(-kW_B)_t+(B-kVW_B)_x=\langle R\rangle,\\
    &(-kW_V)_t+(A-kVW_V)_x=\langle uR\rangle,
    \end{split}
\end{equation}
which can be simplified with the use of Eq.~(\ref{eq12}). Besides that, we introduce
the `long derivative'
\begin{equation}\label{eq20}
    \frac{D}{Dt}=\frac{\prt}{\prt t}+V\frac{\prt}{\prt x}
\end{equation}
and substitute (\ref{eq18}) into (\ref{eq12}) to obtain the complete set of the Whitham
modulation equations:
\begin{equation}\label{eq21}
\begin{split}
  &\frac{DW_A}{Dt}=W_A\frac{\prt V}{\prt x},\\
  &\frac{DW_B}{Dt}=W_A\frac{\prt B}{\prt x}-W_A{\langle R\rangle},\\
  &\frac{DW_V}{Dt}=W_A\frac{\prt A}{\prt x}-W_A{\langle uR\rangle}.
  \end{split}
\end{equation}
Naturally, they differ from the original Whitham equations \cite{whitham-65} only by the terms with
$\langle R\rangle$ and $\langle uR\rangle$.

A remarkable discovery of Whitham was that the unperturbed modulation equations can be transformed
``after considerable manipulation'' to the diagonal (Riemann) form and the Riemann invariants are
expressed in terms of zeros $\al,\,\beta,\ga$.
In our case the same transformation leads again to the
diagonal form of Whitham equations, however now with small right-hand sides, that is the Whitham equations
become non-uniform.
The necessary `considerable manipulation' is described in  detail in \cite{kamch-2000} and we
indicate here briefly the main steps only.

First, we transform Eqs.~(\ref{eq21}) from the variables $A,\,B,\,V$ to the variables $\al,\,\beta,\ga$
with the use of relationships (\ref{eq6}),
\begin{equation}\label{eq22}
    \begin{split}
&    W_{A,\alpha}\frac{D\alpha}{Dt}+W_{A,\beta}\frac{D\beta}{Dt}+W_{A,\gamma}\frac{D\gamma}{Dt}\\
    &=2W_A(\alpha_x+\beta_x+\gamma_x),\\
&    W_{B,\alpha}\frac{D\alpha}{Dt}+W_{B,\beta}\frac{D\beta}{Dt}+W_{B,\gamma}\frac{D\gamma}{Dt}=-W_A\\
&    \times[(\beta+\gamma)\alpha_x+(\alpha+\gamma)\beta_x+(\alpha+\beta)\gamma_x]\\
&-W_A \langle R\rangle ,\\
&    W_{V,\alpha}\frac{D\alpha}{Dt}+W_{V,\beta}\frac{D\beta}{Dt}+W_{V,\gamma}\frac{D\gamma}{Dt}=-W_A\\
&    \times[\beta\gamma\cdot\alpha_x+\alpha\gamma\cdot\beta_x+\alpha\beta\cdot\gamma_x]
-W_A\langle uR\rangle,
    \end{split}
\end{equation}
where
\begin{equation}\label{eq23}
    \begin{split}
    W_{A,\alpha}&=\frac{1}{\sqrt{8}}\oint\frac{du}{(u-\alpha)\sqrt{f(u)}},\\
    W_{B,\alpha}&=-\frac{1}{\sqrt{8}}\oint\frac{udu}{(u-\alpha)\sqrt{f(u)}},\\
    W_{V,\alpha}&=-\frac{1}{\sqrt{8}}\oint\frac{(u^2/2)du}{(u-\alpha)\sqrt{f(u)}},
    \end{split}
\end{equation}
and similar expressions can be written for derivatives with respect to $\beta$ and $\ga$. Next, we multiply the first
equation (\ref{eq22}) by $p=\la\beta+\al\ga-\beta\ga$, the second equation by $q=2\al$, the third equation by $r=2$
and add them; then with the use of the identities
\begin{equation}\nonumber
    \begin{split}
    W_{A,\alpha}&+W_{A,\beta}+W_{A,\gamma}=\frac{1}{\sqrt{8}}\oint\frac{f'(u)du}{f^{3/2}(u)}=0,\\
    pW_{A,\alpha}&+qW_{B,\alpha}+rW_{V,\alpha}\\
    &=-\frac{1}{\sqrt{8}}\oint\frac{d}{du}
\left(2\sqrt{\frac{(u-\beta)(u-\gamma)}{-(u-\alpha)}}\right)du=0
    \end{split}
\end{equation}
we obtain
\begin{equation}\label{eq24}
    \begin{split}
    &\frac{D(\beta+\gamma)}{Dt}+\frac{W_A}{W_{A,\alpha}}\frac{\prt(\beta+\gamma)}{\prt x}\\
    &=-\frac1{(\al-\beta)(\al-\ga)} \frac{W_A}{W_{A,\alpha}}(\al\langle R\rangle-\langle uR\rangle)
    \end{split}
\end{equation}
and similar equation can be obtained for the variables $\al+\beta$ and $\al+\ga$ by means of
cyclic transposition of the parameters $\al,\,\beta,\,\ga$. At last, we introduce the Riemann
invariants of the unperturbed KdV equation,
\begin{equation}\label{eq25}
\begin{split}
  &\lambda_1=-\tfrac12(\alpha+\beta),\quad \lambda_2=-\tfrac12(\alpha+\gamma),\\
  &\lambda_3=-\tfrac12(\beta+\gamma),\quad \la_1\leq\la_2\leq\la_3,
  \end{split}
\end{equation}
and with account of
$$
\frac{W_A}{W_{A,\alpha}}=\frac{2W_A}{W_{A,\lambda_3}}=\frac{2L}{\prt L/\prt\lambda_3}
$$
and similar formulae for $W_A/W_{A,\beta}$ and $W_A/W_{A,\gamma}$ we
arrive at the Whitham equation in the form
\begin{equation}\label{eq26}
\begin{split}
  &\frac{\prt\la_i}{\prt t}+v_i^{(0)}\frac{\prt\la_i}{\prt x}=\frac{L}{\prt L/\prt\lambda_i}
  \frac{\langle(2\la_i-s_1-u) R\rangle }{4\prod_{j\neq i}(\la_i-\la_j)}\\
  & i=1,2,3,
  \end{split}
\end{equation}
where velocities $v_i^{(0)}$ are given by
\begin{equation}\label{eq27}
\begin{split}
  v_i^{(0)}=-2s_1+\frac{2L}{\prt L/\prt\lambda_i},
  \quad i=1,2,3,
  \end{split}
\end{equation}
and $s_1=\la_1+\la_2+\la_3$. Here all the variables should be parameterized by the Riemann invariants
$\la_1,\,\la_2,\,\la_3$.
In particular, the periodic solution of the KdV equation takes the form
\begin{equation}\label{eq28}
\begin{split}
  &u(x,t)=\la_3-\la_2-\la_1\\
  &-2(\la_3-\la_2)\,\sn^2(\sqrt{\la_3-\la_1}\,(x-Vt),m),
  \end{split}
\end{equation}
where
\begin{equation}\label{eq29}
\begin{split}
V&=-2s_1=-2(\la_1+\la_2+\la_3),\\
m&=\frac{\la_3-\la_2}{\la_3-\la_1},
\end{split}
\end{equation}
and the wavelength is given by
\begin{equation}\label{eq30}
  L=\frac1k=\frac{2K(m)}{\sqrt{\la_3-\la_1}}.
\end{equation}
Substitution of (\ref{eq30}) into (\ref{eq27}) gives expressions for the Whitham velocities in their
original form \cite{whitham-65},
\begin{equation}\label{eq31}
  \begin{split}
  v_1^{(0)}&=-2s_1
  +\frac{4(\la_3-\la_1)(1-m)K(m)}{E(m)},\\
  v_2^{(0)}&=-2s_1
  -\frac{4(\la_3-\la_2)(1-m)K(m)}{E(m)-(1-m)K(m)},\\
  v_3^{(0)}&=-2s_1
  +\frac{4(\la_3-\la_2)\K(m)}{E(m)-K(m)},
  \end{split}
\end{equation}
where $E(m)$ is the complete elliptic integral of the second kind.
For averaging the perturbation terms, it is convenient to introduce the variable $\mu=(u+s_1)/2$
which changes within the interval $\la_2\leq\mu\leq\la_3$ and satisfies the equation
\begin{equation}\label{eq32}
    \mu_x=2\sqrt{-P(\mu)},
\end{equation}
where
\begin{equation}\label{eq32b}
\begin{split}
    P(\mu)&=(\mu-\la_1)(\mu-\la_2)(\mu-\la_3)\\
    &=\mu^3-s_1\mu^2+s_2\mu-s_3,
    \end{split}
\end{equation}
consequently
\begin{equation}\label{eq33}
    \langle \mathcal{Q}\rangle=\frac1L\int_0^L \mathcal{Q}dx
    =\frac{1}{L}\int_{\la_2}^{\la_3}\frac{\mathcal{Q}\,d\mu}{\sqrt{-P(\mu)}}.
\end{equation}

These formulae easily reproduce the results found in Refs.~\cite{gp-87,akn-87,gp-91} with $R=\epsilon u_{xx}$
corresponding to the Burgers viscosity and permitted one to derive Whitham equations for the case of
$R=F(t)u-G(t)u^2$ corresponding to shallow water waves over a gradual slope with account of bottom
friction \cite{egk-07}. However, as was mentioned in Introduction, this approach fails if averages of
$R$ and/or $uR$ vanish what happens when these expressions are the $x$-derivatives of some other expressions
and, hence, they contribute into the fluxes of the conservation laws (\ref{eq13}). We shall consider
such a situation in the next subsection.

\subsection{Whitham equations for the case of gradient perturbations}

Here we shall assume that both $R$ and $uR$ are space derivatives
\begin{equation}\label{eq15}
    R=\mathcal{Q}_{1,x},\quad uR=\mathcal{Q}_{2,x}.
\end{equation}
and, hence, they make
additional contribution into the fluxes in the conservation laws (\ref{eq13}).
Then all divergence terms in (\ref{eq13}) should be treated on the same footing and averaging
of the conservation laws yields
\begin{equation}\label{eq16b}
  \begin{split}
&  \langle u\rangle_t+\langle 3u^2+u_{xx}\rangle_x=\langle \mathcal{Q}_{1} \rangle_x,\\
&  \langle \tfrac12u^2\rangle_t+\langle 2u^3+uu_{xx}-\tfrac12u_x^2\rangle_x=\langle \mathcal{Q}_{2} \rangle_x.
  \end{split}
\end{equation}
Transformations similar to those which were done above give instead of Eq.~(\ref{eq21}) the equations
\begin{equation}\label{eq21b }
\begin{split}
  &\frac{DW_A}{Dt}=W_A\frac{\prt V}{\prt x},\\
  &\frac{DW_B}{Dt}=W_A\frac{\prt B}{\prt x}-W_A\frac{\prt \langle\mathcal{Q}_1\rangle}{\prt x},\\
  &\frac{DW_V}{Dt}=W_A\frac{\prt A}{\prt x}-W_A\frac{\prt \langle\mathcal{Q}_2\rangle}{\prt x},
  \end{split}
\end{equation}
or, after transition to the variables $\al,\,\beta,\,\ga$, the equations
\begin{equation}\label{eq22b}
    \begin{split}
&    W_{A,\alpha}\frac{D\alpha}{Dt}+W_{A,\beta}\frac{D\beta}{Dt}+W_{A,\gamma}\frac{D\gamma}{Dt}\\
    &=2W_A(\alpha_x+\beta_x+\gamma_x),\\
&    W_{B,\alpha}\frac{D\alpha}{Dt}+W_{B,\beta}\frac{D\beta}{Dt}+W_{B,\gamma}\frac{D\gamma}{Dt}=-W_A\\
&    \times[(\beta+\gamma)\alpha_x+(\alpha+\gamma)\beta_x+(\alpha+\beta)\gamma_x]\\
&-W_A(\langle\mathcal{Q}_1\rangle_{\al}\alpha_x+\langle\mathcal{Q}_1\rangle_{\beta}\beta_x+\langle\mathcal{Q}_1\rangle_{\ga}\gamma_x),\\
&    W_{V,\alpha}\frac{D\alpha}{Dt}+W_{V,\beta}\frac{D\beta}{Dt}+W_{V,\gamma}\frac{D\gamma}{Dt}\\
&    =-W_A[\beta\gamma\cdot\alpha_x+\alpha\gamma\cdot\beta_x+\alpha\beta\cdot\gamma_x]\\
&-W_A(\langle\mathcal{Q}_2\rangle_{\al}\alpha_x+\langle\mathcal{Q}_2\rangle_{\beta}\beta_x+\langle\mathcal{Q}_2\rangle_{\ga}\gamma_x).
    \end{split}
\end{equation}
Their linear combination used above for transition to Eq.~(\ref{eq24}) now gives
\begin{equation}\label{eq24b}
    \begin{split}
    &\frac{D(\beta+\gamma)}{Dt}+\frac{W_A}{W_{A,\alpha}}\frac{\prt(\beta+\gamma)}{\prt x}\\
    &=-\frac1{(\al-\beta)(\al-\ga)}\frac{W_A}{W_{A,\alpha}}\{(\al\langle\mathcal{Q}_1\rangle_{\al}-\langle\mathcal{Q}_2\rangle_{\al})\al_x\\
    & + (\beta\langle\mathcal{Q}_1\rangle_{\beta}-\langle\mathcal{Q}_2\rangle_{\beta})\beta_x
    +(\ga\langle\mathcal{Q}_1\rangle_{\ga}-\langle\mathcal{Q}_2\rangle_{\ga})\ga_x \}.
    \end{split}
\end{equation}
At last, transformation to the variables $\la_1,\,\la_2,\,\la_3$ (see (\ref{eq25})) yields the Whitham equations in the form
\begin{equation}\label{eq26b}
\begin{split}
  &\frac{\prt\la_i}{\prt t}+v_i^{(0)}\frac{\prt\la_i}{\prt x}=
  \frac1{4\prod_{j\neq i}(\la_i-\la_j)}\frac{L}{\prt L/\prt\lambda_i}\\
  &\times\sum_k\left\{(2\la_i-s_1)
  \frac{\prt\langle \mathcal{Q}_1\rangle}{\prt\la_k}- \frac{\prt\langle\mathcal{Q}_2\rangle}{\prt\la_k}\right\}\frac{\prt\la_k}{\prt x},\\
  & i=1,2,3.
  \end{split}
\end{equation}
Here it was supposed that both $R$ and $uR$ are gradients of `fluxes' $\mathcal{Q}_1$ and $\mathcal{Q}_2$, respectively.
If only one of the variables $R$ and $uR$ can be represented as a gradient, then only one corresponding term $\langle \mathcal{Q}_k\rangle$
is included in (\ref{eq26b}) and the other term must be treated as a non-gradient one resulting in the right-hand sides
of the Whitham equations, as was shown in the preceding subsection.

Generally speaking, the variables $\la_1,\,\la_2,\,\la_3$ in Eqs.~(\ref{eq26b}) are not Riemann invariants anymore.
Indeed, we have got a non-diagonal matrix of Whitham velocities
\begin{equation}\label{eq34}
    \frac{\prt\la_i}{\prt t}+\sum_jv_{ij}\frac{\prt\la_j}{\prt x}=0,\quad v_{ij}=v_i^{(0)}\delta_{ij}+v_{ij}^{(1)},
\end{equation}
where $v_{ij}^{(1)}$ correspond to the perturbation terms. They are much smaller than the contributions $v_i^{(0)}\delta_{ij}$
into the diagonal ones and we can find the characteristic velocities as well as the corresponding eigenvectors in the
way similar to the stationary perturbation theory well known in quantum mechanics (see, e.g., section 38 in \cite{LL-3}) as long as
\begin{equation}\label{eq35}
    |v_i^{(0)}-v_j^{(0)}|\gg \mathrm{max}|v_{ij}^{(1)}|.
\end{equation}
In particular, the characteristic velocities are given in our approximation by the formulae
\begin{equation}\label{eq36}
    v_i\cong v_i^{(0)}+v_{ii}^{(1)}.
\end{equation}
The condition (\ref{eq35}) indicates that Eqs.~(\ref{eq26b}) cannot be applied to situations with degeneration of two
Whitham velocities, that is, for example, to the edge points of dispersive shock waves. This conclusion is confirmed
by the following remark. As is known (see, e.g., \cite{gp-73}), mean value of the variable $u$ vanishes at the soliton
edge of dispersive shock wave according to the law
\begin{equation}\label{eq37}
    \langle u\rangle\propto\frac1{\ln(\la_2-\la_1)}.
\end{equation}
Hence, this mean value has infinite derivatives with respect to Riemann invariants $\la_1$ and $\la_2$, and mean
values $\langle u^n\rangle$ have similar singularities at the soliton edge. Since typically the perturbation terms
are expressed in terms of such mean values and their derivatives, then the characteristic velocities (\ref{eq36}) are
also singular at the soliton edge what prevents application of Eqs.~(\ref{eq26b}) to the theory of DSWs.
A possible method of avoiding this difficulty is discussed in the following subsection.

\subsection{Elimination of gradient perturbations from Whitham equations}

There are situations when the gradient terms can be eliminated from the Whitham equations and the difficulty
indicated in the preceding subsection can be avoided by means of the method used first by Marchant and
Smyth \cite{ms-90} in application of the extended KdV (or Gardner) equation
\begin{equation}\label{eq38}
    u_t+6uu_x+u_{xxx}=6\epsilon u^2u_x
\end{equation}
to formation of undular bores in the resonant flow of a fluid over topography. It is supposed that $\epsilon$
is a small parameter, $\epsilon\ll1$, and Marchant and Smyth showed that with accuracy $O(\epsilon)$ the equation
can be reduced to the KdV equation
\begin{equation}\label{eq39}
    U_t+6UU_x+U_{xxx}=0
\end{equation}
by means of a simple substitution
\begin{equation}\label{eq40}
\begin{split}
    &u=U+\epsilon(U^2+U_{xx}/2)\\
    &\text{or}\\
    &U=u-\epsilon(u^2+u_{xx}/2).
    \end{split}
\end{equation}
In fact, the equation (\ref{eq38}) is completely integrable and the Whitham equations can be derived \cite{pavlov-94,pre-2012}
beyond the perturbation theory for any value (and sign) of $\epsilon$. Nevertheless, as we shall show, the method of
substitutions similar to that of Marchant and Smyth turns out to be quite useful in a situation of the `generalized
KdV equation'
\begin{equation}\label{eq40b}
    u_t+6uu_x+u_{xxx}=6\epsilon F(u)u_x,
\end{equation}
where $F(u)$ is a regular function in the region of variations of $u$.

Let us look at several examples of simple substitutions and the results of the corresponding transformations.
Everywhere we neglect the terms smaller than the order $O(\epsilon)$.

A substitution
\begin{equation}\label{eq41}
    U=u-\epsilon u_{xx}\quad\text{or}\quad u=U+\epsilon U_{xx}
\end{equation}
leads to an approximate (with accepted here accuracy) identity
\begin{equation}\nonumber
    U_t+6UU_x+U_{xxx}=u_t+6uu_x+u_{xxx}+12\epsilon u_xu_{xx}.
\end{equation}
Hence, the perturbed KdV equation
\begin{equation}\label{eq42}
    u_t+6uu_x+u_{xxx}=-12\epsilon u_xu_{xx}
\end{equation}
is reduced to Eq.~(\ref{eq39}) by means of the substitution (\ref{eq41}).

A substitution
\begin{equation}\label{eq43}
    U=u-\epsilon u^2\quad\text{or}\quad u=U+\epsilon U^2
\end{equation}
transforms in a similar way the perturbed KdV equation
\begin{equation}\label{eq44}
    u_t+6uu_x+u_{xxx}=6\epsilon (u^2u_x+u_xu_{xx})
\end{equation}
to Eq.~(\ref{eq39}).
Composition of substitutions (\ref{eq41}) and (\ref{eq43})
yields the substitution (\ref{eq40}) of Marchant and Smyth. In this particular case
the generalized KdV equation (\ref{eq40b}) is reduced to the non-perturbed KdV
equation (\ref{eq39}), however such a reduction is generally impossible. For example,
a substitution
\begin{equation}\label{eq45}
    U=u-\epsilon u^3\quad\text{or}\quad u=U+\epsilon U^3
\end{equation}
reduces the equation
\begin{equation}\label{eq47}
    u_t+6uu_x+u_{xxx}=6\epsilon u^3u_x
\end{equation}
to
\begin{equation}\label{eq48}
    U_t+6UU_x+U_{xxx}=6\epsilon (U_x^3+3UU_xU_{xx}),
\end{equation}
and here the terms in the right-hand side cannot be eliminated by additional substitutions.
In the general case a substitution
\begin{equation}\label{eq49}
    U=u-\epsilon F(u)\quad\text{or}\quad u=U+\epsilon F(U)
\end{equation}
transforms (\ref{eq40b}) to
\begin{equation}\label{eq50}
\begin{split}
    U_t&+6UU_x+U_{xxx}\\
    &=6\epsilon (F^{\prime\prime\prime}(U)U_x^3+3F^{\prime\prime}(U)U_xU_{xx}).
    \end{split}
\end{equation}
Thus, these substitutions transform a perturbed equation to another perturbed equation,
however there is an important difference between the initial and reduced forms: the
perturbation in Eq.~(\ref{eq40b}) is a gradient one whereas in Eq.~(\ref{eq50}) it is
non-gradient and therefore these perturbations should be treated by different types of Whitham equations
discussed above.

Another important feature of the perturbation terms in (\ref{eq50}) is that
they do not contribute into the right-hand sides of the Whitham equations (\ref{eq26})
which coincide, hence, with unperturbed Whitham equations for the KdV equation. Let us demonstrate this
for the averaged value $\langle R\rangle$. Simple integration by parts shows that
$$
\int F^{\prime\prime}U_xU_{xx}dx=-\frac12\int F^{\prime\prime\prime}(U)U_x^3dx
$$
that is contribution of the second term in the right-hand side of Eq.~(\ref{eq50}) reduces after
averaging (up to factor$-1/2$) to the contribution of the first term. But its average vanishes, as
shows a simple calculation (see (\ref{eq4}))
\begin{equation}\nonumber
\begin{split}
\int F^{\prime\prime\prime}(U)U_x^3dx&=\oint F^{\prime\prime\prime}(U)U_x^2dU\\
&= 2\oint F^{\prime\prime\prime}(U)f(U)dU=0
\end{split}
\end{equation}
for regular functions $F(U)$ in the region of variation of $U$. Thus, we can use
unperturbed Whitham equations for Eq.~(\ref{eq50}) and transform the results to Eq.~(\ref{eq40b})
by means of the substitution (\ref{eq49}).

Let us illustrate this approach by its application to the Gurevich-Pitaevskii problem of evolution
of initial step-like pulse
\begin{equation}\label{eq51}
    u(x,0)=\left\{
    \begin{array}{c}
    u_-\quad\text{for}\quad x<0,\\
    0 \quad\text{for}\quad x>0,
    \end{array}
    \right.
\end{equation}
according to Eq.~(\ref{eq40b}). The substitution (\ref{eq49}) transforms this problem to the same
problem for the equation (\ref{eq50}) with $U_-=u_--\epsilon F(u_-)$. The well-known solution
of the Gurevich-Pitaevskii problem for unperturbed Whitham equations applicable to (\ref{eq50})
yields, in particular, the speeds of edges of the dispersive shock wave (see, e.g., \cite{gp-73,kamch-2000}),
\begin{equation}\label{eq52}
    \begin{split}
    s_-&=-6U_-=-6(u_--\epsilon F(u_-)),\\
    s_+&=4U_-=4(u_--\epsilon F(u_-)),
    \end{split}
\end{equation}
where $s_-$ is the speed of the small-amplitude edge and $s_+$ is the speed of the soliton edge.
For $F(u)=u^2$ these formulae coincide with the results of exact theory developed in \cite{pre-2012}
for the Gardner equation and it is instructive to compare them with the results of El's method
\cite{el-05} applied to Eq.~(\ref{eq40b}). According to this method, the speeds of the shock edges
are equal to (see section IV in \cite{el-05})
\begin{equation}\label{eq53}
    s_-=\frac{\prt\om_0}{\prt k}(u_-,k_-),\quad
    s_+=\frac{\widetilde{\om}_s(u_+,\widetilde{k}_s)}{\widetilde{k}_s},
\end{equation}
where $\om_0(u,k)$ and $\widetilde{\om}_s(u,\widetilde{k})$ are the linear and the `soliton' dispersion laws,
respectively, given in our case by
\begin{equation}\label{eq54}
    \begin{split}
    \om_0(u,k)=V(u)k-k^3,\\
    \widetilde{\om}_s(u,\widetilde{k})=V(u)\widetilde{k}+\widetilde{k}^3.
    \end{split}
\end{equation}
and $V(u)=6(u-\epsilon F(u))$. In the problem (\ref{eq51}) we have $u_+=0$ and $k_-$ and $\widetilde{k}_s$
should be found by solving the differential equations
\begin{equation}\label{eq55}
    \begin{split}
    \frac{dk}{du}=\frac{\prt\om_0/\prt u}{V(u)-\prt\om_0/\prt k},\quad k(u_+)=0,\\
    \frac{d\widetilde{k}}{du}=\frac{\prt\widetilde{\om}_s/\prt u}{V(u)-\prt\widetilde{\om}_s/\prt \widetilde{k}},
    \quad \widetilde{k}(u_-)=0.
    \end{split}
\end{equation}
Easy integration yields
\begin{equation}\label{eq56}
    k_-=\widetilde{k}_s=2\sqrt{u_--\epsilon F(u_-)}
\end{equation}
and substitution of these values into (\ref{eq53}) gives Eqs.~(\ref{eq52}).

It is remarkable that the perturbation theory reproduces in this case the results correct beyond applicability
of the perturbation approach. Besides that, in framework of the perturbation theory under consideration
the dissipative effects can be easily taken into account by adding corresponding terms to Eq.~(\ref{eq40b})
and by calculation of their contribution into the right-hand sides of the Whitham equations (\ref{eq26}).
One may suppose that this approach can be useful for consideration of the problems of the type considered
in a recent preprint \cite{ehs-15} where a combined action of nonlinear, dispersive and dissipation effects
should be taken into account.

At last, the perturbation theory is not limited to the step-like initial conditions. For example, if
one wishes to consider evolution after wave breaking described by the Gardner equation (\ref{eq38}), so that the initial
condition can be reduced to the form (see, e.g., \cite{kamch-2000})
\begin{equation}\label{eq57}
    x-6(u-\epsilon u^2)t=-u^3,
\end{equation}
then the substitution (\ref{eq40}) transforms the problem to solving the equation (\ref{eq39}) with the initial condition
\begin{equation}\label{eq58}
    x-6Ut=-U^3-3\epsilon U^4.
\end{equation}
Here the Whitham equations correspond to the unperturbed KdV case. Their solution with the initial
condition (\ref{eq58}) for $\epsilon=0$ was found by Potemin \cite{potemin}, and its generalization to the case
$\epsilon\neq0$ can be done by including higher commuting flows of the KdV hierarchy into
the standard method of integration of unperturbed Whitham equations (see, e.g., \cite{kamch-2000}).

\section{Steady state solutions of perturbed Whitham equations}

In dissipative-dispersive systems one can distinguish several characteristic stages of evolution of a pulse. For example,
if we consider evolution of a step-like initial pulse (\ref{eq51}), then it is natural to expect that the dispersion term
with third order $x$-derivative is much more important at the initial stage of evolution than the dissipative perturbation
term typically proportional to the second order derivative $u_{xx}$ (or even $u$ itself; see, e.g., \cite{egk-07}). Hence,
if we introduce small parameter $R\sim\epsilon$, then for time $t\lesssim 1/\epsilon$ we can neglect the dissipative
perturbation and evolution is described by the classical solution of the Gurevich-Pitaevskii problem \cite{gp-73}.
However, for $t\gtrsim1/\epsilon$ the damping of solitons in the dispersive shock wave train becomes essential and
this damping can be compensated by the non-zero boundary condition at the small amplitude limit $x\to-\infty$
where $u\to u_-$. Thus, for $t\gg1/\epsilon$ we arrive at the steady-state solution of the perturbed KdV equation
with dissipation balanced by the non-zero boundary condition. If damping is small enough, then the steady-state
dispersive shock can be described by a modulated cnoidal wave solution (\ref{eq28}) where $\la_i$ are the functions
of $\xi=x-Vt$ and satisfy the perturbed Whitham equations (\ref{eq26}). They are greatly simplified by noticing
that $s_1=\la_1+\la_2+\la_3$ is their integral $s_1=\mathrm{const}$ provided
$V=-2s_1$ and, hence, the Whitham equations can be reduced to
\begin{equation}\label{eq59}
    \frac{d\la_i}{d\xi}=-\frac{\langle(2\la_i-s_1-u)R\rangle}{8\prod_{j\neq i}(\la_i-\la_j)},
    \quad i=1,2,3.
\end{equation}
It is easy to check that these equations indeed have the integral $s_1=\mathrm{const}$ so that above {\it ansatz} is justified.
Easy calculation shows that two other symmetric functions $s_2=\la_1\la_2+\la_1\la_3+\la_2\la_3$ and $s_3=\la_1\la_2\la_3$
(see (\ref{eq32})) satisfy the equations
\begin{equation}\label{eq60}
    \begin{split}
    \frac{ds_2}{d\xi}=\frac14\langle R\rangle,\quad
    \frac{ds_3}{d\xi}=\frac18[s_1\langle R\rangle+\langle uR\rangle].
    \end{split}
\end{equation}
Thus, we have reduced the problem to the system of two ordinary differential equations for $s_2,\,s_3$ and $\la_i$ are
considered as functions of $s_2,\,s_3$ being the roots of the algebraic equation
\begin{equation}\label{eq61}
    \la^3-s_1\la^2+s_2\la-s_3=0.
\end{equation}

Especially simple and practically important situation realizes when $\langle R\rangle=0$, hence we have the second integral
$s_2=\mathrm{const}$ and the problem reduces to a single differential equation
\begin{equation}\label{eq62}
    \frac{ds_3}{d\xi}=\frac18\langle uR\rangle.
\end{equation}
Instead of introduction of new dependent variable $s_3$ we can return in this case to Eqs.~(\ref{eq59}) and consider,
say, $\la_1$ and $\la_2$ as functions of $\la_3$ where $\la_3=\la_3(\xi)$. Then we get
\begin{equation}\label{eq63}
    \frac{d\la_1}{d\la_3}=\frac{\la_3-\la_2}{\la_2-\la_1},\quad \frac{d\la_1}{d\la_3}=-\frac{\la_3-\la_1}{\la_2-\la_1},
\end{equation}
and this system has, as we know, two integrals
\begin{equation}\label{eq64}
\begin{split}
    &\la_1+\la_2+\la_3=s_1=\mathrm{const},\\
    &\la_1\la_2+\la_1\la_3+\la_2\la_3=s_2=\mathrm{const}.
    \end{split}
\end{equation}
Consequently, $\la_1$ and $\la_2$ as functions of $\la_3$ are two roots of the quadratic equation
\begin{equation}\label{eq65}
    \la^2-(s_1-\la_3)\la+s_2-(s_1-\la_3)\la_3=0.
\end{equation}
They are ordered according to inequality  $\la_1\leq\la_2$ and the integration constants $s_1,\,s_2$ can be found from
the boundary conditions. Substitution of expressions for $\la_1=\la_1(\la_3)$ and $\la_2=\la_2(\la_3)$ into
differential equation for $\la_3(\xi)$ (see (\ref{eq59})) with known $\langle uR\rangle$ solves in principle
the problem. However, some important consequences can be obtained without integration of this differential
equation. Let us illustrate this by application of the derived formulae to the steady-state dispersive shock wave
evolved from the initial step-like distribution (\ref{eq51}).

The initial distribution (\ref{eq51}) suggests that at $x\to-\infty$ the shock has a form of a small-amplitude wave
with $m\to0$, $\la_2\to\la_3$, so that here according to (\ref{eq28}) we have $\la_1=\la_1^-=-u_-$ and the integrals
take the form
\begin{equation}\label{eq66}
    s_1=-u_-+2\la_3^-,\quad s_2=-2u_-\la_3^-+(\la_3^-)^2.
\end{equation}
At the soliton edge $m\to1$, $\la_2\to\la_1$ the wave (\ref{eq28}) reduces to
$$
u(x,t)=-\la_3+\frac{2(\la_3-\la_1)}{\cosh^2[\sqrt{\la_3-\la_1}\,(x-Vt)]}
$$
and since from (\ref{eq51}) we have the boundary condition $u(x,t)\to0$ as $x\to\infty$, we get here
\begin{equation}\label{eq68}
    \la_1^+=\la_2^+,\quad \la_3^+=0.
\end{equation}
Then the leading soliton at the soliton edge is described by the equation
\begin{equation}\label{eq69}
    u(x,t)=\frac{-2\la_1}{\cosh^2[\sqrt{-\la_1}\,(x+4\la_1t)]}.
\end{equation}
Now we substitute the boundary conditions (\ref{eq68}) into Eq.~(\ref{eq65}) to obtain the relation
\begin{equation}\label{eq70}
    s_1^2-4s_2=0
\end{equation}
between the integration constants which must be fulfilled along the whole dispersive shock wave.
Its application to Eqs.~(\ref{eq66}) yields at the small amplitude edge $\la_1^-=-u_-$,
$\la_2^-=\la_3^-=-u_-/4$ and hence
\begin{equation}\label{eq69b}
    s_1=-\frac32u_-,\quad s_2=\frac9{16}u_-^2.
\end{equation}
Then equation (\ref{eq65}) gives at the soliton edge where $\la_3^+=0$ the value
of a double root $\la_1^+=\la_2^+=-3u_-/4$ and as a result we find the expressions for the
soliton amplitude $a_s=-2\la_1$ and the velocity $V=-2s_1$ of the shock wave in terms of a given
value of the initial step amplitude,
\begin{equation}\label{eq71}
    a_s=\frac32u_-,\quad V=3u_-.
\end{equation}
These expressions were obtained long ago by Johnson \cite{johnson-70} for a particular case
of Burgers dissipation $R=\epsilon u_{xx}$ in framework of direct perturbation technique. As we see, they
can be reproduced quite easily by the Whitham method under more general assumptions about the form of
dissipative perturbation of the KdV equation. It is remarkable that under certain conditions not only
the velocity $V=3u_-$ of the shock does not depend on the details of irreversible processes but also the
amplitude of the leading soliton has a `universal' value $a_s=(3/2)u_-$.

\section{Conclusion}

During past fifty years the Whitham theory has developed into a vast branch of applied mathematics with
various applications to real physical processes related with nonlinear wave propagation. In spite of such
a progress, as we have demonstrated in this paper, the original Whitham approach remains quite effective.
Here we have shown its
fruitfulness for a perturbed KdV equation, however it is clear that it is not confined to this single
application. In fact, the main difficulty in the original direct Whitham's approach was the problem of finding 
Riemann invariants and Whitham found these Riemann invariants for the KdV equation case due to clever
insight and skillful
calculations. Now, due to discovered after publication of Whitham's paper relationship between 
the Whitham theory and the finite-gap integration method
of completely integrable equations, the Riemann invariants have been found for many equations which belong to
the AKNS scheme \cite{akns-74} (see, e.g., \cite{kamch-2000}). Therefore we can use the
averaged conservation laws in any parametrization with account of perturbation terms and after that
transform these equations to the known Riemann invariants of the unperturbed system arriving at perturbed
Whitham equations. Thus, one may hope that the direct Whitham method can find in future many
interesting and important applications.

\section*{Acknowledgments}

I am grateful to G.A.~El, R.H.J.~Grimshaw, M.A.~Hoefer, N.~Pavloff, and M.V.~Pavlov for useful discussions of the
Whitham theory.

\end{document}